\documentclass[runningheads]{llncs}
\usepackage{graphicx}
\usepackage{subfigure}
\usepackage{amsfonts}
\usepackage{multirow}
\usepackage[font=scriptsize]{caption}
\usepackage[colorlinks, linkcolor=red,anchorcolor=blue,citecolor=green]{hyperref}
\begin{document}

\title{Domain Adaptive Cascade R-CNN for MItosis DOmain Generalization (MIDOG) Challenge}
\titlerunning{Domain Adaptive Cascade R-CNN for MIDOG Challenge}
\author{Xi Long\inst{1} \and
	Ying Cheng\inst{1} \and
	Xiao Mu\inst{1} \and
	Lian Liu\inst{2} \and
    Jingxin Liu\inst{1}
}

\institute{Histo Pathology Diagnostic Center, Shanghai, China \and
	Department of Electrical and Computer Engineering,  Technical University of Munich, Germany \\
	\email{jingxin.liu@outlook.com}}
\maketitle              % typeset the header of the contribution

\begin{abstract}
% Deep convolutional neural network based recognition and segmentation rely on the identical distributed training data and its corresponding pixel-level labelled ground truth. Since collecting and labelling medical data are widely regarded as tedious and time consuming processes, the algorithms that able to recover the performance of the pre-trained network to new unlabelled dataset are highly desired. In this paper, domain adaptation with dual adversarial learning modules is utilized to adapt the deep segmentation network trained on source dataset to target dataset domain. In particular, semantic adaptation is to shift the source image domain to the target image domain; while structural adaptation is used considering the spatial similarities between two different domains. We evaluate our new approach on two pairs of datasets of MRI and pathological images. We show that the proposed method outperforms other state-of-the-art models.

We present a summary of domain adaptive cascade R-CNN method for mitosis detection of digital histopathology images. By comprehensive data augmentation and adapting existing popular detection architecture, our proposed method has achieved an F1 score of 0.7500 on the preliminary test set in MItosis DOmain Generalization (MIDOG) Challenge at MICCAI 2021.

\end{abstract}
\section{Introduction}
Mitotic count (MC) is a common and critical marker of breast cancer prognosis \cite{elston1991pathological}. Manually marking mitotic cells in Hematoxylin and Eosin (H\&E) stained histopathology images is obviously time-consuming and subjective. With the dramatic improvements in computer vision and digital pathology, researchers proposed to automate this process in pathology laboratories. A number of mitosis detection competitions have been held, e.g., the ICPR MITOS-2012 challenge \cite{roux2013mitosis}, the ICPR MITOS-ATYPIA-2014 challenge \cite{roux2014mitos}, and the MICCAI-TUPAC16 challenge \cite{veta2019predicting}. Thus, numerous works have been proposed, and achieved remarkable success in the field of mitosis detection \cite{bertram2019large,sebai2020mask}. 

However, deep learning based detection models may have poor generalization capability to unseen datasets due to the domain shift. Such problem is commonly observed in digital histopathology image analysis, caused by tissue preparation and image acquisition. The MItosis DOmain Generalization
(MIDOG) challenge \cite{marc_aubreville_2021_4573978}, hosted as a satellite event of the 24$^{th}$ International Conference at Medical Image Computing and Computer Assisted Intervention (MICCAI) 2021, addresses this topic in the form of assessing MC on a multiscanner dataset. In this abstract, we propose a method with domain augmentation and Domain Adaptive Cascade R-CNN (DAC R-CNN) for mitosis detection to achieve robust detection performance for varieties of images. 

\section{Materials}
The MIDOG training set consists of 200 image tiles from Whole Slide Images (WSIs) of human breast cancer tissues with H\&E dye. The image tiles were digitized with four slide scanners: Hamamatsu XR nanozoomer 2.0, Hamamatsu S360 (0.5 NA), Aperio ScanScope CS2, and Leica GT450, resulting in 50 image tiles per scanner. From each image tile, a trained pathologist selected an area of 2$mm^2$ corresponding to approximately 10 high power fields. Annotations are provided for the first three scanners.

\section{Methodology}

\subsection{Stain Color Domain Augmentation}
In MIDOG challenge, the test set contains images scanned by unknown slide scanners. Previous stain normalization methods transfer different pathology images into one target stain color style, which may not improve the robustness of detection model when dealing with unseen stain color appearance. Therefore, in addition to traditional image augmentation methods, we propose stain color domain augmentation to generate training images with a wider range of stain color appearances, making our model more robust to unseen data. To be specific, we build on previous stain normalization methods by adding randomness in selecting normalization methods and target color styles. 

The proposed stain color domain augmentation method involves two types of stain normalization methods: \emph{Reinhard} \cite{reinhard2001color} and \emph{Vahadane} \cite{vahadane2015structure}. Each method will be executed with a given probability. \emph{Reinhard} transfers color based on target mean and variance, while \emph{Vahadane} transfers color according to the target color appearance matrix. We obtain an initial range for target mean, variance, and each element of the color appearance matrix using the whole training set, respectively. Target values are randomly selected from those ranges during augmentation, making it possible to generate images with very different color styles. We will gradually enlarge those ranges to create new training samples to feed the network until detection performance degrades to a limit. In this way, we expect the trained network to achieve robust detection performance for varieties of images.

\subsection{Domain Adaptive Cascade R-CNN}
\begin{figure}[t]
	\centering
	\includegraphics[width = 8cm]{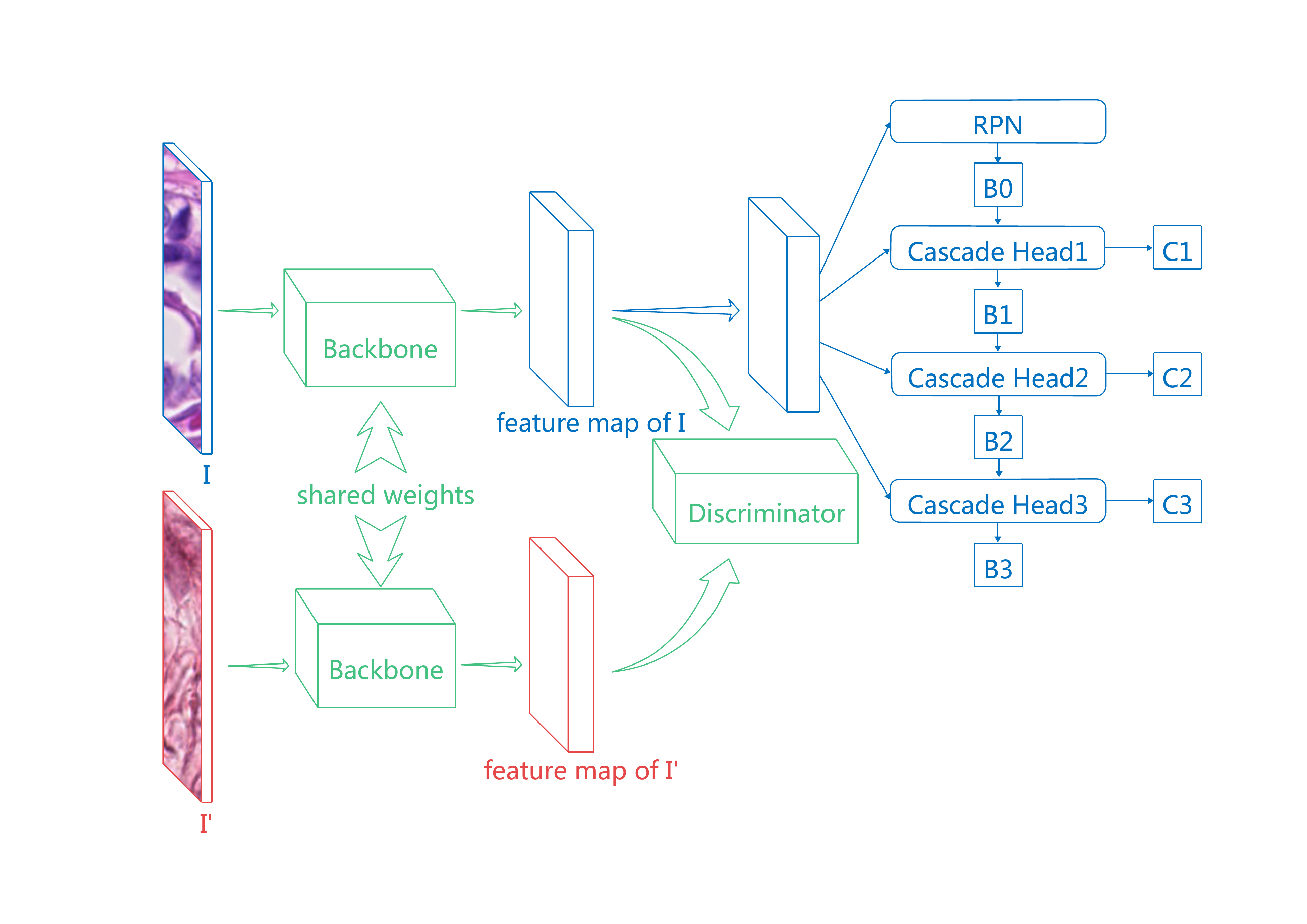} 
	\caption{Domain Adaptive Cascade R-CNN architecture. ``I'' is input image,  ``I$'$'' reference image for domain adaptation, ``Backbone'' backbone network for feature map extraction, ``Cascade Head" detection head, ``B'' bounding box, and ``C'' classification. ``Discriminator'' is a convolutional PatchGAN classifier for distinguishing input image from reference image, which should be removed in inference phase. } 
	\label{fig1}
\end{figure}

We propose a Cascade R-CNN based domain adaptation model for mitosis detection \cite{cai2019cascade}, referred to as Domain Adaptive Cascade R-CNN (DAC R-CNN) (See Fig.\ref{fig1}). The backbone network of DAC R-CNN is pre-trained ResNet-50, and three cascaded detection heads are utilized for high quality detector. Inspired by our previous work \cite{hou2019dual}, we employ an image-level adaptation component to address overall differences between different image domains like image color and style using PatchGAN\cite{isola2017image}, through which we aim to obtain similar feature maps from input image and reference image. 

Specifically, a parallel reference branch is added to the network architecture with a discriminator following feature maps of both the input image and the reference image. Two branches share the same backbone, which serves as the generator. The discriminator distinguishes the input image from the reference image. The discriminator is a convolutional PatchGAN classifier that operates on image patches. One advantage of PatchGAN is that it can be applied to images with arbitrary sizes. Note that labels of reference images are not needed. By generating similar features, domain adaptation seeks to achieve comparable performance for unlabelled data.

\section{Results}
Our proposed method produced an F1 score of 0.7500 with a 0.7792 precision and a 0.7229 recall on the preliminary test set in MItosis DOmain Generalization (MIDOG) Challenge at MICCAI 2021. 

\small
\bibliographystyle{splncs04}
\bibliography{mybib}

\end{document}